\documentclass{article}
\usepackage{amsmath}
\usepackage[super,sort&compress,comma]{natbib}
\usepackage{hyperref}
\usepackage{orcidlink}
\usepackage{doi}
\usepackage[margin=2cm]{geometry}
\usepackage{authblk}

\title{\textbf{Continuous thermochemical sources of AlF molecules}}
\date{}

\author[1,*]{Pulkit Kukreja~\orcidlink{0009-0000-3061-6154}} 
\author[1,*]{Priyansh Agarwal~\orcidlink{0009-0009-3490-8619}} 
\author[1]{Maximilian Doppelbauer~\orcidlink{0000-0002-6288-0256}}
\author[1]{Jionghao Cai~\orcidlink{0009-0004-8263-6834}} 
\author[1]{Xiangyue Liu~\orcidlink{0000-0003-4463-8068}} 
\author[1]{Eduardo Padilla~\orcidlink{0000-0001-8569-6593}} 
\author[1]{Sebastian Kray~\orcidlink{0000-0002-8599-3800}}
\author[1]{Henrik Haak}
\author{Russell Thomas~\orcidlink{0000-0003-4348-7368}}
\author[2]{Stefan Truppe \orcidlink{0000-0002-0121-6538}} 
\author[1]{Boris G. Sartakov~\orcidlink{0000-0001-9498-7587}}
\author[1]{Gerard Meijer~\orcidlink{0000-0001-9669-8340}} 
\author[1,$\dagger$]{Sid Wright~\orcidlink{0000-0003-2431-5624}}

\affil[1]{\textit{Fritz Haber Institute of the Max Planck Society, Faradayweg 4-6, 14195 Berlin, Germany}}

\affil[2]{\textit{Blackett Laboratory, Imperial College London, London SW7 2AZ, United Kingdom}}

\begin{document}

\maketitle
\def\thefootnote{*}\footnotetext{These authors contributed equally to this work}
\def\thefootnote{$\dagger$}\footnotetext{sidwright@fhi-berlin.mpg.de}
\vspace{-0.7cm}

\begin{abstract}
The AlF molecule, currently subject to laser cooling and trapping efforts, has the advantage that it can be efficiently produced in a thermochemical reaction between sublimated aluminum trifluoride and aluminum metal. Here we present a series of experiments with continuous molecular beam sources of AlF, utilising this reaction. We demonstrate a compact AlF molecular beam oven whose total far-field brightness is $5\times 10^{15}$ molecules per steradian per second at 923~K, just below the melting temperature of aluminum metal. The continuous output from the oven begins to exceed the peak brightness of a jet-cooled, ablation-based supersonic AlF source for the $v=0$, $J=7$ level, and we obtain an excellent signal-to-noise ratio with the oven in pulsed laser ionisation spectroscopy experiments. By delivering flux from the oven into a cryogenic Ne buffer gas cell, we lower the rotational temperature of the beam to around 30~K and reduce its most probable forward velocity from 600~ms$^{-1}$ to 200~ms$^{-1}$. In addition, we demonstrate that AlF can be made in a simple dispenser package, and observe that molecules thermalise to the laboratory temperature after colliding with vacuum chamber walls of the experiment. The resulting transient AlF vapour may enable direct loading of a molecular magneto-optical trap.
\end{abstract}

\section{Introduction}

Aluminum monofluoride (AlF), a deeply bound diatomic molecule with a $^1\Sigma^+$ electronic ground state, has recently become the first spin-singlet molecule to be laser-cooled and loaded into a magneto-optical trap (MOT)\cite{Padilla2025}. AlF has a large binding energy of 6.9~eV\cite{Gross1954}, an ionisation potential of 9.7~eV\cite{Walter2024}, and a negative electron affinity of about $-0.84$~eV \cite{Behera2025}. In combination, these properties make it chemically stable compared to the spin-doublet molecules that have been laser-cooled thus far. The $A^1\Pi\leftarrow{} X^1\Sigma^+$ deep ultraviolet transition in AlF is highly vibrationally diagonal, and the short radiative lifetime of the $A^1\Pi$ state ($\tau = 1.9$~ns \cite{Truppe2019}) enables large radiation pressure forces for laser cooling. Similar to CO \cite{Gilijamse2007}, the lowest energy spin-triplet state in AlF, the $a^3\Pi$ state, is metastable, and decays exclusively to the ground state on millisecond timescales\cite{Walter2022}. Moreover, the $a^3\Pi \rightarrow{}X^1\Sigma^+$ decay is also highly vibrationally diagonal\cite{Truppe2019}, offering the potential for narrow line laser cooling and precision spectroscopy in ultracold molecular samples. These features position AlF as a unique system within the field of cold molecules, and are a strong motivation for future experiments. 

At present, molecular MOTs \cite{barry_magneto-optical_2014, truppe_molecules_2017, anderegg_radio_2017,collopy_3d_2018,vilas_magneto-optical_2022,zeng_three-dimensional_2024,lasner_magneto-optical_2025} are loaded by first generating the target species in the gas phase via laser ablation, often involving a chemical reaction driven at thousands of degrees Kelvin. The molecules are then cooled by collisions with helium buffer gas in a cryogenic ($\sim4$~K) environment \cite{Hutzler2012}, resulting in intense pulses of molecules that are slow enough for laser-deceleration and trapping. However, these sources are large, relatively complex and expensive, and prone to day-to-day fluctuations in molecular beam properties. Traps are typically loaded with repetition rates of 1 or 2~Hz to maintain a slow and cold molecular beam, and effective removal of the buffer gas in the source is essential to ensure the required vacuum quality at the trap location. These drawbacks are accepted when laser cooling spin-doublet molecules, that are typically difficult to generate in the gas phase without ablation, and for which laser cooling is simplest when exciting the $N=1$ rotational level (generally requiring source temperatures below 10~K). 

Certain properties of AlF, resulting from its distinct electronic structure and chemistry, favour exploring alternative molecular beam sources for trapping experiments. Firstly, AlF can be produced efficiently in the gas phase, via the endothermic reaction,
\begin{equation}
    \mathrm{AlF_3(g)} + 2\mathrm{Al(s/l)} \rightarrow 3\mathrm{AlF(g)} \hspace{0.3cm}. 
    \label{eqn:chemAlF}
\end{equation}

\noindent The thermochemistry of this reaction is well understood, and results in a $p$-$T$ relation for AlF that is similar to atomic Ca. Second, the electronic structure of AlF, in particular its $A^1\Pi\leftarrow{}~X^1\Sigma^+$ transition, enables straightforward laser cooling for any excited rotational level via closed $Q(J)$ rotational lines. This somewhat relaxes the need for cryogenic cooling in the molecular beam source. Thirdly, since only the excited level of this transition has a significant Zeeman effect, the design of a Zeeman slower is relatively simple for AlF. This method of laser-slowing \cite{Phillips1982}, being time-independent, is ideal for accumulating molecules from continuous sources, and widely used in atomic trapping experiments.

In this study, we present various molecular beam sources of AlF based on reaction \eqref{eqn:chemAlF}, and characterise these sources with laser cooling and trapping in mind. We demonstrate an intense, continuous molecular beam with a far-field beam brightness of $5\times10^{15}$~sr$^{-1}$s$^{-1}$ at $T=923$~K, just below the melting temperature of aluminum metal (section \ref{sec:MBE}). Using this source, we improve the spectroscopic constants of $v=0-4$ levels of the $A^1\Pi$ state via laser-induced fluorescence spectroscopy, and study high-lying rovibrational levels of the $c^3\Sigma^+$ state for the first time, via pulsed laser ionisation spectroscopy (section \ref{sec:cState}). In a proof-of-principle experiment, we cool the continuous AlF beam by collisions with cold Ne in a buffer gas cell (section \ref{sec:BGcooling}). The rotational temperature of the beam is reduced to around 30~K and the most probable forward velocity reduces to around 200~ms$^{-1}$, at which point laser slowing is practically feasible. Finally, we demonstrate that atomic dispensers can be adapted to produce AlF, and observe that molecules survive collisions with the vacuum chamber walls to produce a transient room temperature vapour (section \ref{sec:Dispenser}). This may enable direct loading of an AlF magneto-optical trap in a compact, cryogen-free setup.

\section{Experimental}

\subsection{Thermochemistry}

\begin{figure}[!tb]
\centering
\includegraphics[width=0.6 \linewidth,trim = {0 0.3cm 0 0}]{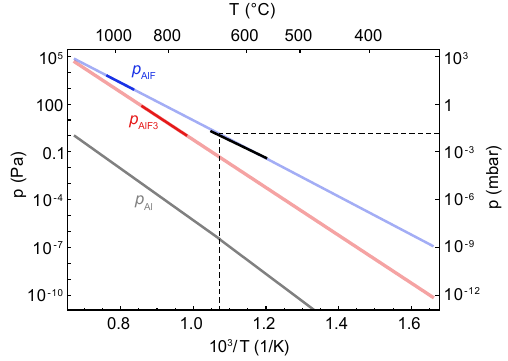}
\caption{Vapour pressures of AlF (transparent blue) and AlF$_3$ (transparent red) versus temperature. For convenience, the upper $x$-axis gives the temperature in $^{\circ}$C, and the right hand $y$-axis gives the pressure in mbar. Solid lines indicate the measurement range in Ko et al.\cite{Ko1965} (red, blue) and in Witt and Barrow\cite{Witt1959} (black). The grey line shows the vapour pressure of aluminum \cite{Alcock1984} and the dashed, black line indicates $p_{\mathrm{AlF}}$ at the melting point of aluminum metal.}
\label{fig:1}
\end{figure}

Thermodynamic properties of reaction \eqref{eqn:chemAlF} were clarified in 1965 by Ko and co-workers \cite{Ko1965}, resolving discrepancies in the earlier literature. Using the transpiration technique, they measured the sublimation pressure of AlF$_3$, $p_{\mathrm{AlF_3}}$, and the equilibrium constant $K(T)$ of reaction \eqref{eqn:chemAlF}. Of interest to us is the equilibrium vapour pressure of AlF, $p_{\mathrm{AlF}}$, versus temperature, which is obtained via the formula,

\begin{equation}
p_{\mathrm{AlF}} = [K(T) p_{\mathrm{AlF_3}}]^{1/3}  \hspace{0.3cm}.  
\end{equation}

\noindent Figure \ref{fig:1} plots $p_{\mathrm{AlF}}$ and $p_{\mathrm{AlF_3}}$ versus $1/T$ using the parameters of Ref. \cite{Ko1965}, and includes the results of Witt and Barrow who measured $p_{\mathrm{AlF}}$ at lower temperatures, near the melting point of aluminum ($933$~K) \cite{Witt1959}. At 973~K ($700$~$^{\circ}$C), $p_{\mathrm{AlF}}=4\times10^{-2}$~mbar, and for all lower temperatures, $p_{\mathrm{AlF}}$ exceeds $p_{\mathrm{AlF_3}}$ by more than an order of magnitude. The vapour pressure of atomic aluminum, $p_{\mathrm{Al}}$ shown in grey in Figure \ref{fig:1}, is negligible in comparison. Together, this permits producing AlF as the dominant component in a molecular beam at moderate source temperatures. We note that at ambient laboratory temperatures ($\sim$300~K), $p_{\mathrm{AlF}}$ is orders of magnitude below the vacuum level of our experiments ($10^{-7} - 10^{-8}$~mbar).

\subsection{Continuous molecular beam source}
\label{sec:MBE}
\begin{figure*}[!tb]
\centering
    \includegraphics[width = 1.0\textwidth, trim = {0 0.4cm 0 0}]{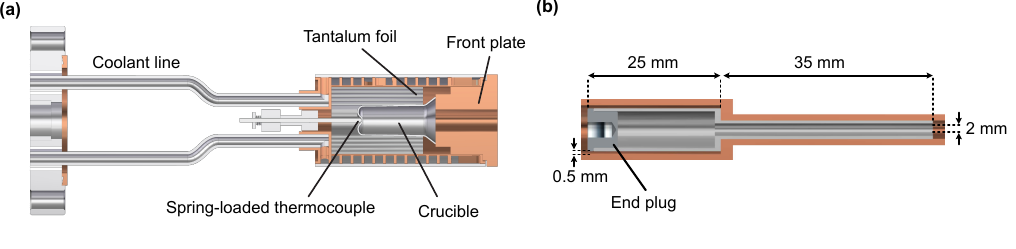}
\caption{(a) Schematic of the Knudsen effusion thermochemical AlF oven. (b) Capillary-capsule which is inserted into the crucible, and holds the Al and AlF$_3$ reagents. The outer copper layer contacts the stainless steel reagent holder throughout the capillary section, but is separated by a 0.5~mm gap along the reagent capsule, ensuring the stainless steel is hottest along the capillary section to prevent clogging.}
\label{fig:2}
\end{figure*}

Figure \ref{fig:2}a shows a diagram of our molecular beam oven, which is a variation of the ultrahigh vacuum Knudsen effusion cell presented in Ref. \cite{Shukla2004}. We use a pyrolytic boron nitride crucible as a heater vessel, surrounded by tantalum wire heater elements (diameter 0.4~mm) along its length. The external body of the cell is water-cooled to reduce the radiation heat load into the vacuum chamber. Five layers of tantalum foil (thickness 0.1~mm), spaced with twisted tungsten wire, are placed between the heater elements and the water-cooled surfaces. This provides sufficient thermal isolation for efficient heating. A spring-loaded K-type thermocouple gently presses against the back of the crucible, and provides the temperature reading for a feedback loop. 

The reagents are held within a capillary capsule construction that slots into the crucible, shown in Figure \ref{fig:2}b. It is composed of an internal stainless steel reagent holder welded to a 35~mm long capillary channel, a stainless steel end plug, and an external copper holder. The copper holder directly contacts the stainless steel along the complete length of the capillary, but is separated by a 0.5~mm gap along the reagent capsule. This heats the capsule through the capillary via contact with the copper, preventing clogging, whilst avoiding reaction between Al and Cu at the operating temperatures used. We separately tested a modified copper/steel capillary capsule with a 90$^{\circ}$ bend of radius 3~cm extending outside the effusion cell, and obtained stable signal with this method. The copper front plate, mounted to the water-cooled exterior of the cell, secures the crucible against the spring-loaded thermocouple and reduces the radiation heat load to the surrounding vacuum chamber. A stainless steel plug slots into the rear of the reagent capsule to prevent material escaping from the back, and can be easily removed to load fresh reagents. For most operating temperatures in this study, the mean free path for AlF molecules is larger than the length of the capillary. In this regime, the capillary reduces the total exit rate of AlF molecules from the oven, with a minimal reduction of the on-axis molecular beam brightness \cite{Jones1969,Jones1971}.

We load the oven with a 0.15-0.2~g parcel of aluminum trifluoride crystals (AlF$_3$, Carl ROTH) wrapped in aluminum foil, and find that a stable and bright signal can be obtained above about 770~K, well below the melting temperature of aluminum. A small discontinuity appears in the AlF output at around 933~K (660~$^{\circ}$C). We attribute this to the foil melting, and infer that the temperature of the reagents matches the value set in the feedback loop to within a few~K. We verified that loading the oven with both YbF$_3$/Al and AlF$_3$/Yb mixtures, used in the past to generate YbF at around 1500~K \cite{Hudson2002}, produces an AlF beam of comparable brightness to that obtained with AlF$_3$/Al in our range of operating temperatures. Similar but less reliable results were obtained with Ca/AlF$_3$; it is reasonable to assume that AlF will be produced in abundance when producing monofluoride radicals with reactions involving aluminum \cite{Blue1963, Knight1971}.

To characterise our molecular beam source,  we excite the $A^1\Pi, v' \leftarrow{} X^1\Sigma^+, v''=v'$ vibrationally diagonal bands near 227.5~nm and record laser-induced fluorescence (LIF) spectra. For brevity, we hereafter use the shorthand notation $v\Delta J(J'')$ to refer to rotational lines of vibrationally diagonal bands, e.g. $2P(J'')$ refers to the $A^1\Pi, v'= 2, J''-1 \leftarrow X^1\Sigma^+, v''=2, J''$ rotational line. The molecular beam interacts with continuous-wave light from a frequency-quadrupled Ti:Sa laser ($<2$~MHz linewidth), 70~cm downstream of the capillary exit. An 8~mm aperture at the entrance of the detector restricts the transverse velocity spread of the molecules to about $\pm 5$~ms$^{-1}$, which slightly broadens the lines beyond their natural linewidth, $\Gamma/(2\pi) = 1/(2\pi \tau)=84$~MHz. The laser light is linearly polarised, and is aligned with a pair of irises mounted either side of the vacuum chamber such that it intersects the molecular beam perpendicularly. Fluorescence is collected with $4f$-optics and imaged onto a photomultiplier tube (PMT), operated in current mode. The fundamental frequency of the Ti:Sa laser (near 910~nm) is monitored via a High-Finesse WS8-10 wavemeter. The wavemeter is calibrated with a temperature stabilised He-Ne laser, providing an absolute accuracy of 60~MHz at 227.5~nm.

\begin{figure*}[tb]
\centering
    \includegraphics[width = 1.0\textwidth,trim = {0 0.4cm 0 0}]{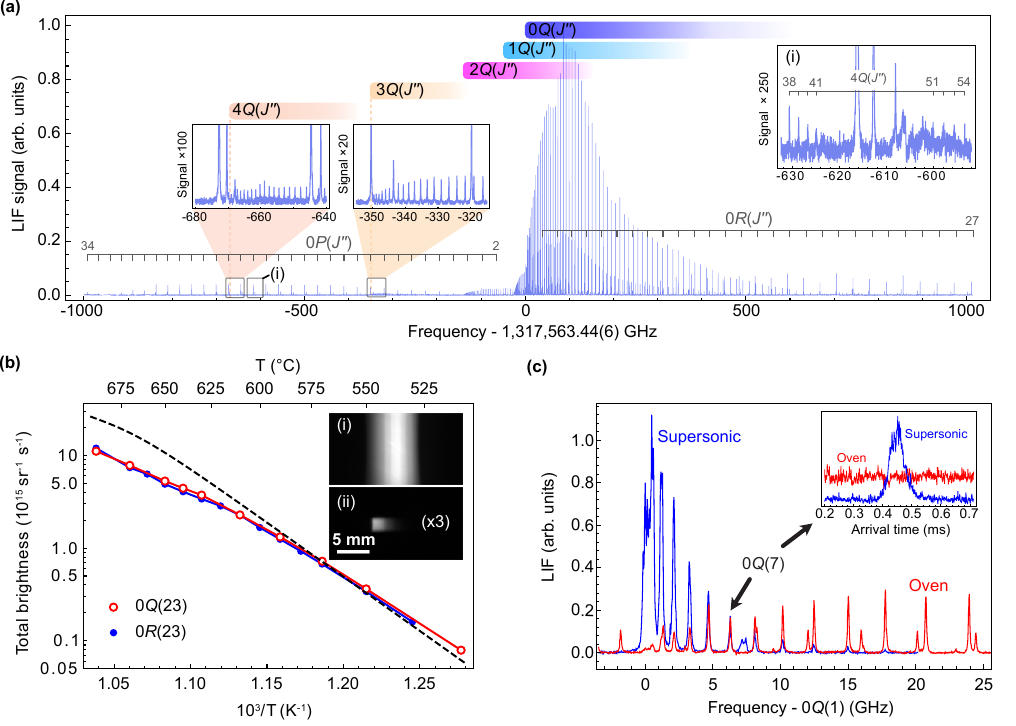}
\caption{(a) Laser-induced fluorescence spectrum of the molecular beam measured 70~cm downstream of the capillary exit, probed on diagonal bands of the \textit{A}$^1\Pi \leftarrow $\textit{X}$^1\Sigma^+$ transition near 227.5~nm. Zoom-ins: close ups of the $3$\textit{Q}$(J'')$ and $4$\textit{Q}$(J'')$ bandheads. Inset (i) shows the region around the 4\textit{Q}(38-54) lines, in which a subset of lines are absent (see text). (b) Total molecular beam brightness, measured using fluorescence on the 0\textit{Q}(23) and 0\textit{R}(23) lines as discussed in the text. The source-detector distance is 42~cm. Dashed, black line: predicted on-axis brightness as discussed in the text. Insets: camera images of (i) Rayleigh scattered light from N$_2$ gas in the detector (1 atm), using 15~mW probe light; (ii) LIF from the AlF molecular beam when saturating the 0\textit{R}(23) line, with the oven temperature set to 873~K. (c) Direct comparison of laser-induced fluorescence spectra of the thermochemical and supersonic AlF beams, in the same setup. The oven set temperature is 903~K.}
\label{fig:3}
\end{figure*}

Figure \ref{fig:3}a shows a spectrum obtained by scanning over a $\pm1$~THz frequency range around the strong $0Q(J'')$ branch, with the oven temperature set to 923~K. This spectrum was recorded over a period of one week without reloading reagents, scanning the laser frequency at a rate of about 20~MHz per second. It is composed of several 50-150~GHz wide segments; between each segment we adjusted the second harmonic conversion modules in the laser system, to ensure an optical power of between 2-3~mW at the experiment, in a few mm diameter beam. We assigned 523 rotational lines within vibrationally diagonal bands up to and including $v=4$, and fit these with a Gaussian lineshape to find their centre frequencies. The $Q$-branches are closed under angular momentum and parity selection rules, and molecules excited on the $0Q(J'')$ lines may scatter an average of 213(30) photons before optical pumping to $v''=1$ in the $X^1\Sigma^+$ state \cite{Hofsaess2021} occurs. The $R(J''$)- and $P(J''+2$) lines connect to the same $e$-symmetry rotational level in the $A^1\Pi$ state. Molecules excited on these lines are optically pumped after emitting an average of three photons or fewer, with the precise value determined by the relevant H{\"o}nl-London factors \cite{Hofsaess2021}. As a result, these lines have correspondingly lower intensities in Figure \ref{fig:3}a. In inset (i) of figure \ref{fig:3}a, we show a portion of the spectrum around the $4Q(38$-$54)$ lines, in which nine $4Q(J'')$ lines are missing. This is due to crossing of the $A^1\Pi,v=4$ and $b^3\Sigma^+,v=3$ levels, which perturb each other via the spin-orbit interaction as discussed in Ref \cite{Barrow1974}. Optical cycling on the $vQ(J'')$ lines is highly sensitive to electronic state mixing, and when the interaction is near-resonant, $A^1\Pi\rightarrow X^1\Sigma^+$ fluorescence from the molecular beam reduces substantially. Hyperfine structure in the lines\cite{Truppe2019} is only resolvable when exciting to levels with $J'=1$ or 2, and is dominated by the structure of the $A^1\Pi$ levels. At larger values of $J'$ it simply serves to broaden the lineshape; for the $0Q(50)$ line, the total hyperfine structure spans 27~MHz, roughly one third of the natural linewidth of the transition. 

Energies of the $X^1\Sigma^+, v=0-8$ levels in AlF have been measured by Bernath and co-workers by high accuracy infrared emission spectroscopy \cite{Zhang1995}. Using these values and the fitted line centres from Figure \ref{fig:3}a, we determine the energies of the $A^1\Pi$ levels, and fit them to the equation, 
\begin{equation}
\begin{split}
    E_v(J) = &T_v + (B_v \pm q_v/2)J(J+1) \\
    &- D_v[J(J+1)]^2 + H_v[J(J+1)]^3 \hspace{0.3cm},
    \label{eqn:A1Pienergies}
\end{split}
\end{equation}
\noindent Here, $T_v$ is the term energy of vibrational level $v$, $B_v$ is the rotational constant, and the $\pm q_v/2$ term accounts for $\Lambda$-doubling into $e/f$-symmetry levels. $D_v$ and $H_v$ are centrifugal distortion coefficients. Our improved constants are presented in table \ref{tab:APiConsts}. For the $v'=4$ levels, we fit the data including a spin-orbit perturbation interaction between the $A^1\Pi,v'=4$ and $b^3\Sigma^+,v=3$ levels, following the analysis in \cite{Walter2022a} \footnote{We use the perturbation interaction coefficient from Ref. \cite{Barrow1974}, $H_0 = 0.31$~cm$^{-1}$, and note that $\xi$ from Ref. \cite{Walter2022a} is related to $H_0$ by $\xi = \sqrt{2}H_0$.}; we also provide a set of effective rotational constants for the $v'=4$ levels, that accurately describe the measurements using equation \eqref{eqn:A1Pienergies} up to the $v'=4,$~$J'=41$ level. Since only $f$-symmetry levels were measured, we have extrapolated the $\Lambda$-doubling constant $q_4$ using the constants from $v'=0-3$. The root-mean-square difference between the fitted and observed energies is between 14 and 18~MHz for each of $v'=0$-$4$. Our observed lines agree with those of Barrow and co-workers \cite{Barrow1974} within their estimated uncertainty, but are systematically about 1.5~GHz higher in frequency. 

\begin{table*}[tb]
  \centering
  \begin{tabular}{ccccccc}
    \hline
    $v$  & $T_v-T_0$ & $B_v$ & $10^{5}D_v$ &  $10^{10}H_v$ & $10^{3}q_v$ \\
    \hline
$0 $&$    0           $&$   16.601744(20,290) $&$    3.20630(60,1720)$&$   -0.2528(5,88)  $&$   -2.933(3,19) $  \\
$1 $&$ 23738.737(10,20)  $&$   16.437918(30,260) $&$    3.2303(10,220)  $&$   -0.289(10,140)   $&$  -2.884(3,14)       $  \\
$2 $&$ 47103.861(10,20)  $&$   16.272317(30,280) $&$    3.2586(10,290)  $&$   -0.339(20,200)   $&$  -2.855(3,13)       $  \\
$3 $&$ 70088.181(20,50)  $&$   16.104612(40,520) $&$    3.2934(30,670)  $&$   -0.389(50,660)   $&$  -2.795(1,3)        $ \\
$4' $&$ 92682.803(30,150) $&$    15.93119(20,330) $&$    3.094(20,790)  $&$   -9.50(90,1730)    $&$  -2.75^a          $  \\
$4$ & $92682.876(50,170)$ & $15.93233(10,270)$ & $3.258(10,400)$ &$-1.34(20,450)$ & $-2.75^a$ \\
    \hline
  \end{tabular}
  \caption{\label{tab:APiConsts} Table of spectroscopic constants for the $A^1\Pi$ state of AlF, obtained by fitting the fluorescence spectrum in Figure \ref{fig:3}a to equation \eqref{eqn:A1Pienergies}. All values are given in GHz, and $T_0=1,329,553.52(6)$~GHz is referenced to the minimum of the $X^1\Sigma^+$ internuclear potential. Values in brackets are the standard deviation in the last digit (SD) and the standard deviation multiplied by the $\sqrt{Q}$, where $Q$ is $Q$-factor of the fit. The latter value accounts for parameter correlations and is a better estimate of the true uncertainty \cite{WATSON1977}$^{\dagger}$. The first row of constants for the $v'=4$ levels are effective constants valid up to $J'=41$, and the second row includes the spin-orbit perturbation interaction with the $b^3\Sigma^+, v=3$ levels (see text).}
\end{table*}

To estimate the molecular beam brightness, we compare the optical power scattered in the detector through laser-induced fluorescence, $\mathcal{P}_{\mathrm{AlF}}$, to the signal from Rayleigh-scattered laser light when the same detector is filled with 1~bar of N$_2$ gas at room temperature, $\mathcal{P}_{\mathrm{N}_2}$. The total Rayleigh-scattered power follows the Beer-Lambert law, and the power scattered over propagation length $z$ is,

\begin{equation}
\mathcal{P}_{\mathrm{N}_2} = \mathcal{P}_{\mathrm{in}}\sigma_{\mathrm{N}_2}N_{\mathrm{N}_2}  z \hspace{0.3 cm}.
\label{eq:ScatteredPower}
\end{equation}

\noindent Here, $\mathcal{P}_{\mathrm{in}}$ is the input power, $\sigma_{\mathrm{\mathrm{N}_2}}$ is the isotropically averaged scattering cross-section, $N_{\mathrm{N}_2}$ is the nitrogen density, and $\mathcal{P}_{\mathrm{N}_2}\ll\mathcal{P}_{\mathrm{in}}$. The detected power is then $\mathcal{P}_d = \alpha \mathcal{P}_{\mathrm{N}_2}$, with $0<\alpha<1$ representing the product of the geometric collection efficiency with the transmission of all relevant optics. We use a circular detection lens, with the polarisation of the laser light forming the `magic' angle $\theta_{\mathrm{m}} = \arccos{\frac{1}{\sqrt{3}}}$ with the detection direction, so that we are insensitive to anisotropy in the emission pattern of the scattered light. Since the density of nitrogen and its Rayleigh scattering cross-section \cite{Ityaksov2008} are known, equation \eqref{eq:ScatteredPower} enables calibrating $\mathcal{P}_{\mathrm{AlF}}$ against $\mathcal{P}_{N_2}$, i.e. it allows measurement of the molecular beam brightness without calculating or estimating $\alpha$. 

We measure the beam brightness for the $v''=0,J''=23$ level, using two methods. In one method, we excite the $0Q(23)$ line with low intensity light, so that we can use equation \eqref{eq:ScatteredPower} with the two-level resonant photon scattering cross-section, $\sigma_{\mathrm{AlF}} = \frac{\lambda^2}{2\pi} = 8.24 \times 10^{-11}$~cm$^2$. This enables calculation of $N_{\mathrm{AlF}}$, and assuming a Maxwell-Boltzmann velocity distribution for the molecular beam, the molecular beam brightness. Alternatively, we excite the $0R(23)$ line, and fully saturate the fluorescence, in which case molecules in the $v''=0,J''=23$ level are optically pumped into the $v=0, J''=25$ level, after scattering on average 2.042 photons. In this case $\mathcal{P}_{\mathrm{AlF}}$ is determined by the rate at which molecules in the $v''=0,J''=23$ level enter the detector, $\dot{n}$. The scattered power is simply $\mathcal{P_{\mathrm{AlF}}} = hc/\lambda \times 2.042\;\dot{n}$, and by using a well-defined molecular beam aperture, we determine $\dot{n}$ and thereby the beam brightness. 

The result of these measurements is shown in Figure \ref{fig:3}b, which plots the total beam brightness (i.e. summed over all rovibrational levels) against the oven set temperature for $550<T<700$~$^\circ$C ($823<T<973$~K). The source-detector distance for these measurements is 42~cm and the molecular beam is restricted by a 2~mm square aperture at the entrance of the fluorescence detector. The data derived from the $0Q(23)$ and $0R(23)$ measurements agree to within about 10 percent. Camera images in the inset show the Rayleigh scattered light (inset (i), with an incident optical power of 15~mW) and $0R(23)$ fluorescence (inset (ii), with the power increased to 30~mW), for an oven temperature of 873~K. AlF molecules travel from left to right, and are optically pumped in the first few mm of interaction with the probe light, such that the fluorescence is fully saturated. At $T=923$~K (just below the melting temperature of aluminum metal), the beam brightness in the $v''=0,J''=23$ level is $6.2\times 10^{13}$~sr$^{-1}$s$^{-1}$. The total beam brightness here is $5\times 10^{15}$~sr$^{-1}$s$^{-1}$, and we estimate that the reagents are consumed at a rate of around 1~mg per hour. The predicted source brightness is shown in Figure \ref{fig:3}b as a dashed, black line. Here, we assume the equilibrium vapour pressure of AlF within the reagent capsule, and include the effect of the capillary transmission following equation (8) of Ref \cite{Jones1969}. At about $923$~K, the expected mean free path for AlF molecules in the capillary becomes comparable to the channel length, and this begins to reduce the predicted on-axis brightness with respect to that of a thin-walled exit aperture. Given the approximations involved, the model describes data remarkably well, matching the observed brightness within a factor three over the range of temperatures used.

We compared the output of the thermochemical source to that of a pulsed, jet-cooled supersonic beam of AlF used in several previous spectroscopic studies \cite{Truppe2019, Doppelbauer2021, Walter2022, Walter2022a, Walter2024}. Both sources were sequentially installed in the same beam machine, and LIF spectra were recorded at a distance 35~cm away, by exciting $A^1\Pi \leftarrow{}X^1\Sigma^+,$~$0Q(J'')$ lines with cw laser light. The results are shown in Figure \ref{fig:3}c. For the Ne-seeded supersonic beam, with a forward velocity of around 800~ms$^{-1}$ and a rotational temperature of about 10~K, the maximum population is in the $v''=0,J''=2$ level. The signal from the thermochemical beam grows for higher $0Q(J'')$ lines and eventually exceeds that from the supersonic beam. The inset to Figure \ref{fig:3}c shows fluorescence signal from the two sources when exciting the $0Q(7)$ line, which is an isolated line for which both sources have appreciable signal. The useful source output within pulsed detection schemes can be considered to cross at this point. It should however be noted that the much narrower velocity distribution of the supersonic beam remains advantageous in certain circumstances.
\footnotetext{In our notation, $Q$ for parameter $\lambda$ is the product of the corresponding diagonal matrix elements of the matrix $A$ and its inverse $A^{-1}$; $Q$ = $A_{\lambda\lambda}\cdot (A^{-1})_{\lambda\lambda}$, as given in Ref. \cite{WATSON1977}. \\$^a$ The value of $q_4$ has been extrapolated using constants from the $v=0$-$3$ levels.}

\subsection{Ionisation spectroscopy via the $c^3\Sigma^+\leftarrow{} a^3\Pi, v=0-8$ bands}
\label{sec:cState} 
Having compared the supersonic and thermochemical AlF beams, we then decided to combine them in a spectroscopic study of the $c^3\Sigma^+$ state. This state is interesting for a number of reasons. The potential energy minimum of the $c^3\Sigma^+$ state lies in the range of the first dissociation limit of the molecule. \textit{Ab initio} quantum chemical calculations \cite{Wells2011} predict that the density of electronic states becomes rather dense in this region, and that there is a barrier in the $c^3\Sigma^+$ state internuclear potential some $8,000-10,000$~cm$^{-1}$ above its minimum value, outside which the molecule dissociates. Depending on the nature of the barrier, the $c^3\Sigma^+$ levels could be used as a route to populating highly excited vibrational levels in the $X^1\Sigma^+$ state of AlF, via distributed vibrational band intensities. However, since levels in the $c^3\Sigma^+$ state have only been observed up to $v=3$ \cite{Barrow1974}, the barrier height can easily be uncertain to a few thousand cm$^{-1}$. 

\begin{figure*}[!tb]
\centering
    \includegraphics[width = \textwidth,trim={0 0.3cm 0 0}]{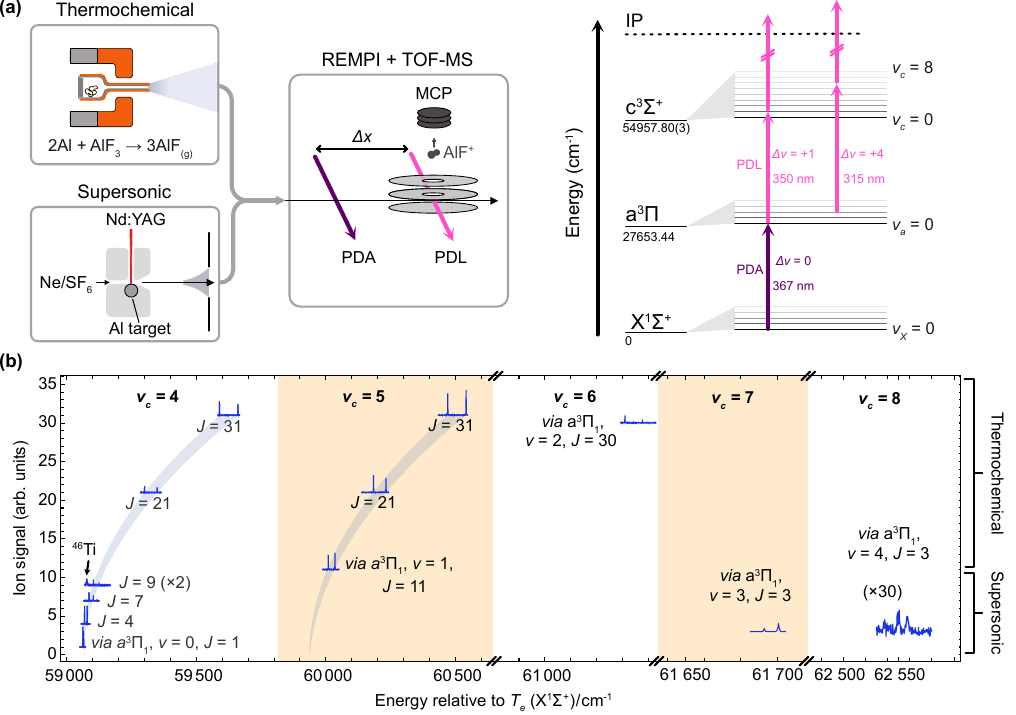}
\caption{REMPI spectroscopy of the \textit{c}$^3\Sigma^+$ state of AlF. (a) Schematic of the experimental setups (left) and energy level diagram showing the relevant transitions (right, relative to the minimum of the \textit{X}$^1\Sigma^+$ state internuclear potential). AlF molecules are prepared in the \textit{a}$^3\Pi$ levels using light from a pulsed dye amplifier (PDA) and subsequently ionised in a 1+1 REMPI process via the \textit{c}$^3\Sigma^+$ state using light from a pulsed dye laser (PDL). The AlF$^+$ cations are mass-selectively detected with a microchannel plate (MCP). (b) 1+1 REMPI spectra of the \textit{c}$^3\Sigma^+,$~$v=4$-$8$ levels. Each spectrum is vertically offset by the starting value of $J$ in the \textit{a}$^3\Pi_1$ state, and shown on an absolute energy scale relative to the potential minimum of the $X^1\Sigma^+$ state. The vibrational level in the \textit{c}$^3\Sigma^+$ state is labelled at the top in each region of the plot. Low-lying levels are measured using the supersonic source and high-lying levels are measured using the thermochemical source, as indicated on the right. The oven set temperature is $903$~K.}
\label{fig:4}
\end{figure*}

To investigate the $c^3\Sigma^+$ state in more detail, we applied resonance-enhanced multi-photon ionisation (REMPI) spectroscopy. The level scheme and experimental setup are shown in Figure \ref{fig:4}a. AlF molecules from the supersonic and continuous thermochemical sources enter a Wiley-McLaren time-of-flight mass spectrometer (TOF-MS), in which single-color, two-photon ionisation via the $c^3\Sigma^+ \leftarrow a^3\Pi$ transition is performed. To operate both sources simultaneously, we mount them coaxially to produce counterpropagating molecular beams, with each a distance $35$~cm from the center of the detector. 

The experiment proceeds as follows. Molecules are first excited to a specific rotational level of the $a^3\Pi_1$ manifold via vibrationally diagonal bands of the $a^3\Pi \leftarrow $~$X^1\Sigma^+$ transition. The excitation light near 367~nm is produced by frequency doubling pulses from a pulsed dye amplifier (PDA, 8~mJ pulse energy, 5~ns pulse duration), injection-seeded with narrow-band cw Ti:Sa laser light near 733~nm. We use the $R_2(J)$ lines of this transition, each of which populates a single rotational level of $e$-symmetry in the $a^3\Pi$ state. The molecules then travel a distance $\Delta x$ in the metastable state before interacting with light from a pulsed dye laser (PDL, frequency doubled to the  310-370~nm wavelength range, pulse energy 2-10~mJ in a few mm beam diameter), that excites the $c^3\Sigma^+ \leftarrow{}a^3\Pi$ transition. The pulse duration and frequency bandwidth of the PDL light are $5$~ns and $0.1$~cm$^{-1}$ respectively, sufficient to generate AlF$^+$ cations when tuned to excite a rotational line of the $c^3\Sigma^+\leftarrow{}a^3\Pi$ transition. The narrow velocity distribution in the supersonic beam allows preparation in a downstream vacuum chamber, and we use $\Delta x= 16$~cm; this is reduced to $\Delta x = 1.2$~cm with the thermochemical beam to reduce the spatial extent of the prepared molecules at the location of the PDL beam, ensuring a good spatial overlap with the ionisation light. 

To study the $v=0-3$ levels of the $c^3\Sigma^+$ state, we excited vibrational bands with $v_c= v_a$ and $v_c= v_a+1$, which have large Franck-Condon factors (above 0.1). To reach the $v=4-8$ levels we excited bands with $v_c= v_a+4$. The Franck-Condon factors for these are estimated to be between $10^{-3}$ and $10^{-4}$, and increasing with $v_a$. The photon energies of the PDA and PDL light are used to determine the excitation energy of the $c^3\Sigma^+$ levels above the initial level in the $X^1\Sigma^+$ state. This results in an absolute energy uncertainty of about $0.1$~cm$^{-1}$, dominated by reproducibility of the PDL frequency measurement across different days. Uncertainty in the frequency of the PDA light is more than an order of magnitude smaller, as are residual Doppler shifts due to misalignment of the laser and molecular beams.

In Figure \ref{fig:4}b, we show overview spectra targeting newly observed rovibrational levels in the $c^3\Sigma^+$ state, i.e. those levels with $v_c=4$-$8$. Spectra are shown with a $y$-axis offset proportional to the value of $J$ prepared in the $a^3\Pi$ state; those with $J>10$ are measured using the thermochemical source.  The $x$-axis energy scale is referenced to the minimum of the $X^1\Sigma^+$ internuclear potential, using the values reported by Bernath et al. \cite{Zhang1995}. From each $e$-level in the $a^3\Pi$ state, two rotational levels in the $c^3\Sigma^+$ state can be reached, where the rotational quantum number $N_c = J_a \pm 1$. The full-width at half-maximum of the lines is 0.35~cm$^{-1}$ and limited by the PDL bandwidth in combination with unresolved fine and hyperfine structure of the $c^3\Sigma^+$ state. The thermochemical beam provides access to highly excited levels, and the signal-to-noise ratio attainable is around 80. This is similar to that obtained with the supersonic beam under the same averaging conditions. Measurements using the thermochemical source are immune from contaminating ion signals that arise due to ablation in the supersonic source. An example of this can be seen in the spectrum starting from the $a^3\Pi_1,v=0,J=9$ level in Figure \ref{fig:4}b, where a spurious ion signal appears at the low frequency side. This arises from (1+1) one-color REMPI of the $3d^2(4s^2)$~$^3F_2 \rightarrow{}3d^3(^4F)4p$~$^3G_3$ transition in atomic Ti, which is produced in the ablation process and then ionised by the PDL light. The $^{46}$Ti isotope (relative natural abundance $8\%$) appears coincident with the AlF$^+$ cations in time-of-flight measurements.

We fit the energies of 49 measured levels of the $c^3\Sigma^+$ state using a Dunham expansion, 

\begin{equation}
    E(v_c,N_c) = \sum_{ n,m =0} Y_{mn} (v_c+1/2)^m \times [N_c(N_c+1)]^n \hspace{0.1cm},
    \label{eq:cState}
\end{equation}

\noindent where the summation runs up to $n+m \leq2$. The results are provided in table \ref{tab:cState}, where we provide the uncorrelated and correlated uncertainty estimates, $\delta Y_{mn}$ and $\sqrt{Q} \delta Y_{mn}$, respectively. The root mean square deviation of the data from the fit is 0.11~cm$^{-1}$, in good agreement with our absolute energy uncertainty. Moreover, extending the Dunham expansion up to $n+m \leq 3$ introduced new parameters that were all zero within their respective $\sqrt{Q} \delta Y_{mn}$ values, and increased the overall uncertainty of the original coefficients. We therefore conclude that the $c^3\Sigma^+$ state can be considered regular for at least 1~eV above its internuclear potential minimum, contrary to expectations. There is no evidence of the onset of a barrier, and the spectroscopic constants of Barrow et al. \cite{Barrow1974} are sufficient to reproduce all measured energies to within 1~cm$^{-1}$.

\begin{table*}
\small
  \begin{tabular*}{\textwidth}{@{\extracolsep{\fill}}lllllll}
    \hline
    Coefficient & Value & $\delta Y_{mn}$ & $\sqrt{Q}*\delta Y_{mn}$ & Barrow \cite{Barrow1974} \\
    \hline
    $Y_{00}$ & 54957.613  & 0.0096 & 0.0353 & $T_e=54957.72\pm 3$\\
 $Y_{10}$ & ~~~933.573  & 0.0020 & 0.0186 & $\omega_e=933.66\pm 3$\\
 $Y_{20}$ & ~~~-\,4.79689  & 0.00036 & 0.00237 & $\omega_ex_e=4.81\pm 1$\\
 $Y_{01}$ &~~~~~~ 0.588798  & 0.000025 & 0.000248 & $B_e=0.58861\pm 4$\\
 $Y_{11}*10^3$ & ~~~\,-4.48794  & 0.0048 & 0.0349 & $\alpha_e\cdot 10^3=4.57\pm 2$\\
 $Y_{02}*10^6$ &  ~~~\,-1.32757  & 0.026 & 0.144 &\\
    \hline
  \end{tabular*}
  \caption{Fitted Dunham expansion coefficients for the \textit{c}$^3\Sigma^+$ state of AlF using equation \eqref{eq:cState}. All values are given in cm$^{-1}$, and $Y_{00}$ is given relative to the minimum of the $X^1\Sigma^+$ state internuclear potential.}
  \label{tab:cState}
\end{table*}

\subsection{Buffer gas cooling}
\label{sec:BGcooling}
Loading AlF molecules from the thermochemical source into a MOT requires that an appreciable fraction of molecules in a single rotational level are within the capture velocity for laser slowing. A promising route to cold, high intensity continuous beams has been demonstrated by seeding a supersonic cryogenic He beam with atoms from an effusive oven source\cite{Huntingdon2023}, but this is rather challenging. 
A number of studies have been published on buffer gas cooling of continuous sources in cryogenic setups\cite{Patterson2009,Patterson2015,Singh2018,Porterfield2019,DiSarno2019}, but in most cases the target species is probed inside or near the exit of the cell, or guided from outside the cell to remove them from the residual buffer gas \cite{vanBuuren2009}. Ablation-based buffer gas beams can be operated at high repetition rates to produce cold, quasi-continuous beams \cite{Shaw2020}, although this comes at the expense of interrupted operation and accelerated target degradation.  

\begin{figure*}[tb]
\centering
    \includegraphics[width = 1.0\textwidth, trim = {0 0.3cm 0 0}]{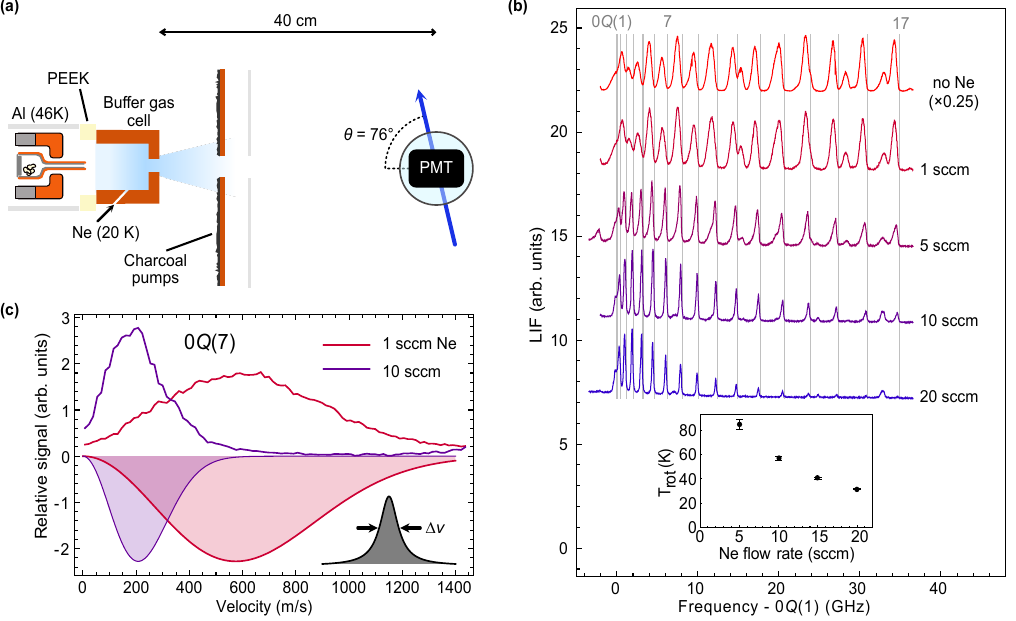}
\caption{Buffer gas cooling of the thermochemical AlF beam. (a) A schematic view of the setup showing oven, buffer gas cell and cryogenic system. A 2~cm PEEK spacer isolates the buffer gas cell from the water-cooled body of the oven, and its black-body radiation is shielded by an external aluminum tube connected to the first cryocooler stage. Ne gas is flowed into the cell at 20~K to cool the AlF molecules, which exit through a circular aperture of diameter 4~mm. Doppler-sensitive laser-induced fluorescence is collected by exciting the \textit{X}$^1\Sigma^+ \leftarrow $\textit{A}$^1\Pi$ transition with a cw laser 40~cm downstream from the cell exit. The laser intersects the molecular beam at the angle $\theta = 76(1)^{\circ}$. (b) Doppler-sensitive LIF spectra around the 0\textit{Q}($J$) bandhead showing rotational and translational cooling as the flow rate of Ne gas into the cell is increased. Grey sticks show the position of the 0\textit{Q}(1-17) resonances. (c) Velocity distributions deduced from the 0\textit{Q}(7) lineshape, for Ne flow rates of 1~sccm and 10~sccm. Shaded curves pointing downwards are Maxwell-Boltzmann velocity distributions for AlF at temperatures of 900~K (red) and 120~K (blue). The shaded black curve at the bottom right indicates the velocity  resolution $\Delta$\textit{v} resulting from the transition natural linewidth and the intersection angle $\theta$.}
\label{fig:5}
\end{figure*}

We implemented buffer gas cooling of the thermochemical AlF beam, with the aim of generating a cold, continuous molecular source suitable for trapping experiments. 
A schematic of the setup used is shown in Figure \ref{fig:5}a, and is adapted from the one presented in Ref. \cite{Wright2022}. The water-cooled external body of the oven is thermally isolated from the cell by a 1~cm section of polyethylene ether ketone (PEEK), and its black-body radiation is shielded from the second stage of the cold head by an aluminum tube contacting the first stage of the cryocooler (40~K). The buffer gas cell has an internal bore of diameter 18~mm, a length of 35~mm and a circular exit aperture of diameter 4~mm. A controlled flow of neon (Ne) gas ($\sim20$~K) enters the cell from the side at a 45$^{\circ{}}$ angle to the molecular beam axis, and is pumped outside the cell by surfaces coated with activated coconut charcoal. At a distance of 40~cm downstream of the cell exit, and 30~cm from the radiation shields of the buffer gas source, the AlF molecules interact with a low intensity probe laser, which excites Doppler-sensitive fluorescence on the $A^1\Pi\leftarrow{} X^1\Sigma^+$ transition. The detuning from a particular rotational line, $\delta f$, can be converted to a detected velocity via the Doppler shift formula, $v = \lambda \delta f/\cos\theta$, with $\theta = 76(1)^{\circ}$ the intersection angle between the laser and molecular beams. The conversion factor here from detuning to velocity is 1~ms$^{-1}$ per MHz, which enables simultaneous measurement of the velocity distribution and rotational temperature of the beam without excessive overlap between neighbouring lines. In this configuration, we detect the density within a given velocity window in the detector, rather than the flux entering it.  

Figure \ref{fig:5}b shows a series of spectra around the $0Q(J)$ bandhead for different flow rates of Ne into the cell. The spectra are shown offset, with the upper trace (measured without Ne flow) scaled down by a factor 4. A series of partially overlapping $0Q(J)$ and $1Q(J)$ lines are visible here, but focussing on the $0Q(7)$ or $0Q(15)$ lines that are well separated results in a most probable forward velocity of the beam of 600~ms$^{-1}$. Introducing 1~sccm of Ne into the cell (resulting in a buffer gas density of about $9\times10^{14}$~cm$^{-3}$) does not significantly alter the velocity or rotational distribution downstream but significantly reduces the signal amplitude. We attribute this to a combination of the large transverse velocity component of the Ne flow into the cell and/or cooling of the capillary by the buffer gas, reducing the output flux from the oven. Rotational cooling is clearly visible as the Ne flow is further increased, concentrating population in the low $v, J$ levels. The fitted rotational temperature $T_{\mathrm{rot}}$ is shown versus the Ne flow rate in the inset to panel (b), and reduces to 31.8(5)~K with a Ne flow rate of 20~sccm, close to the temperature measured on the cell body. At the same time, the Doppler shift and broadening of all lines is reduced, as translational cooling occurs.

Velocity distributions extracted from the $0Q(7)$ lineshape are shown in Figure \ref{fig:5}c. The most probable forward velocity of the beam reduces to 200~ms$^{-1}$ when increasing the Ne flow rate from 1~sccm to 10~sccm. Shown as shaded curves pointing downwards are the Maxwell-Boltzmann velocity distribution functions for $T_{\mathrm{trans}} = $900~K (red) and 120~K (blue). The velocity resolution is limited by the large natural linewidth and relatively large intersection angle of the laser and molecular beams; the associated broadening function is shown in the bottom right-hand side of the panel in black. 

We were also able to excite fluorescence directly outside the buffer gas cell, where the AlF density is large, and collect the emitted light using a PMT at the opposite end of the machine, some 85~cm away. A comparison of the $0Q(2)$ signal using a Ne flow rate of 20~sccm, to that with the Ne flow removed, revealed that the overall AlF density outside the cell in low-$J$ levels barely changes after buffer gas cooling, i.e. the cooling process is rather lossy. The even larger reduction in the signal observed downstream, of about a factor four, implies loss of molecules due to increased transverse divergence of the buffer gas cooled beam. Nevertheless, about a factor two more $v=0,J=7$ molecules enter the detector with velocities below 150~ms$^{-1}$ when a Ne flow rate of 10~sccm is used, relative to that without Ne flow; for molecules in the $v=0,J=1$ level, the gain is a factor seven. We expect that improving the design of the buffer gas inlet and thermal isolation of the oven from the cryogenic cell will improve the beam brightness of the cold beam, but leave this to future work. Attempts to cool using similar flow rates of He and D$_2$ buffer gas proved ineffective; we speculate that this is due to the closer mass-matching between AlF and Ne, and the lower thermal conductivity of Ne that enables a hotter capillary.

\subsection{Dispenser source}
\label{sec:Dispenser}
Alkali atom dispenser sources \cite{Wieman1995,Fortagh1998} are ubiquitous in the field of cold and ultracold atoms, and offer a compact, inexpensive and safe means to produce ambient temperature vapour for loading atomic traps. No equivalent source exists for a laser-coolable molecule at present, but this approach is viable for AlF since it can be produced from solid/vapour phase reagents. Moreover, it is worthwhile for species with $^1\Pi \leftarrow{}^1\Sigma^+$ laser cooling transitions, for which multiple excited rotational levels can be feasibly laser-cooled. With this motivation, we have implemented a dispenser source for AlF. We load AlF$_3$ crystals and aluminum foil into a standard atomic dispenser (AlfaVakuo e.U., type 5C). The dispenser has an internal diameter of 5~mm and a length of 70~mm, and is resistively heated using 7-10~A (5-10~W); an indium foil seal is unnecessary since the reagents do not require protection from ambient air. 

\begin{figure*}[tb]
\centering
    \includegraphics[width = 1.0\textwidth,trim = {0 0.3cm 0 0}]{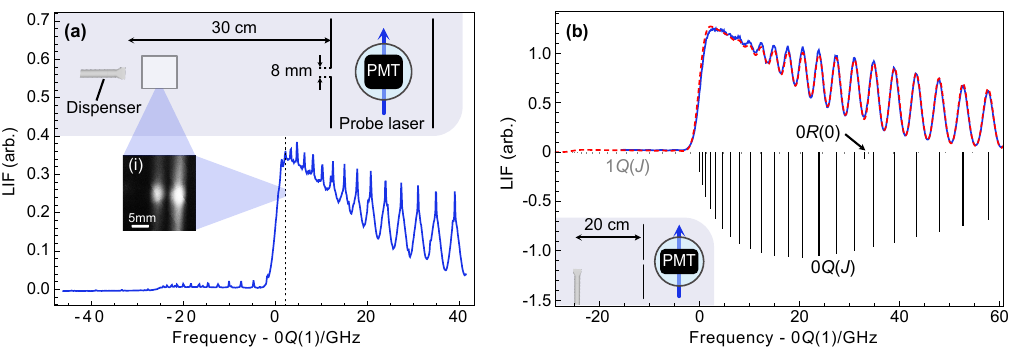}
\caption{Dispenser source for AlF. (a) Laser-induced fluorescence spectrum measured 30~cm from the dispenser exit. The signal is composed of a forward-travelling molecular beam component (narrow peaks), and Doppler-broadened thermal vapour spectrum (broad background features). Inset (i) shows a camera image of the LIF a few cm outside the dispenser taken at the $0Q(4)$ resonance, in which these two signal components are visible. (b) An LIF spectrum taken with the direct line of sight between the dispenser and detector blocked (illustrated by the diagram in the bottom left corner), showing only the Doppler-broadened vapour signal. Dashed, red line: fit to the data as discussed in the text. Black (grey) bars pointing downwards show the intensities of the rotational lines with $v=0$ ($v=1$), assuming a temperature $T_{\mathrm{rot}} = 298$~K.}
\label{fig:6}
\end{figure*}

We first operated the dispenser similar to an atomic beam oven, directing the output flux towards a fluorescence detector located 30~cm downstream. Figure \ref{fig:6}a shows a spectrum after exciting the $A^1\Pi \leftarrow{} X^1\Sigma^+$ transition around the $0Q(J)$ bandhead, using low intensity probe light (1~mW/cm$^2$). A narrow peak appears at each $Q(J)$ resonance whose width is around the natural linewidth, with a broader feature underneath with a full-width at half-maximum of about 2.4~GHz. The inset to the figure shows an image of the fluorescence taken using a UV-sensitive camera a few cm from the dispenser exit, with two parallel probe laser beams exciting the $0Q(4)$ resonance. Two components are visible in the fluorescence image: an intense molecular beam component directed forward, and a broad background component with no apparent preferred direction. We attribute the two components of the fluorescence image to the two signal components in Figure \ref{fig:6}a, and infer that a significant fraction of the AlF molecules emitted by the dispenser thermalise with the room temperature walls of the vacuum chamber before eventually sticking. 

To test this assertion, we mounted the dispenser with its axis perpendicular to the line of sight into the detector, separating the two by 20~cm. We then record another fluorescence spectrum, correcting for variations in the laser intensity with a laser power meter to accurately determine the rotational temperature. As can be seen in Figure \ref{fig:6}b, only the Doppler-broadened component of the signal remains. We fit these data to the distribution, 

\begin{equation}
\begin{split}
S(T_{\mathrm{rot}},T_{\mathrm{trans}}) = \sum_J (2J+1)\exp{[\frac{-E_{v,J}}{(k_BT_{\mathrm{rot}})}]} \\
\times r_{v\mathcal{L}(J)} \exp{[\frac{-m\lambda^2(f-f_{v\mathcal{L}(J)})^2}{(2k_BT_{\mathrm{trans}})}]} \hspace{0.3cm},
\end{split}
\end{equation}

\noindent to determine the rotational and translational temperatures of the vapour, $T_{\mathrm{rot}}$ and $T_{\mathrm{trans}}$ respectively. Here, $E_{v,J}$ are the rovibrational energies of the $X^1\Sigma^+$ levels, and $v\mathcal{L}(J)$ are the rotational lines; $m$ is the molecular mass, $\lambda$ is the transition wavelength, $f-f_{v\mathcal{L}(J)}$ is the detuning of the laser from the line $v\mathcal{L}_{J}$, and $r_{v\mathcal{L}(J))}$ is its branching fraction\footnote{We here neglect vibrational branching, meaning $r
_{0R(0)} = 2/3$, and $r = 1$ for all $vQ(J)$ lines.}. We fix the centres of all $v\mathcal{L}(J)$ falling within the scan range, and fit only $T_{\mathrm{trans}}$, $T_{\mathrm{rot}}$, a common pre-factor for the line intensities, and a $y$-axis offset. The fit returns $T_{\mathrm{rot}} = 275$~K and $T_{\mathrm{trans}} = 297$~K, with statistical error bars of a few Kelvin, confirming that the AlF molecules are well thermalised with the ambient temperature vacuum walls. Individual $0Q(J)$ and $0R(0)$ (1$Q(J)$) line intensities are shown pointing downwards in the figure as black (grey) sticks. The $v=0,J=12$ level has the highest population, although we note that the $v=0,J=1$ level has $\sim20\%$ of this value. In a separate setup, we estimated the vapour density in the $v=0,J=4$ level to be $2.2\times 10^{5}$~cm$^{-3}$. Assuming a spherical capture volume of diameter 1~cm for a magneto-optical trap operating on the $0Q(4)$ line, and a capture velocity of 50~ms$^{-1}$, then the predicted loading rate is around $10^{6}$~s$^{-1}$. Therefore, provided the trap lifetime is in the tens of millisecond range, a few times $10^4$ molecules could be loaded, comparable to that loaded at present from a pulsed, cryogenic molecular beam.

\section{Discussion}
Thermochemical production of AlF compares favourably to that of most laser-coolable molecules. Generating a continuous beam of SrF \cite{Tu2009} with a similar far-field brightness requires temperatures of around 1500~K, and consumes reagents at a rate of 0.2~g per hour. To produce yttrium monoxide (arguably the more chemically stable of the laser-cooled spin-doublet molecules) at the necessary vapour pressures, temperatures in excess of 2500~K are needed \cite{Ackermann1964}. The convenience of reaction \eqref{eqn:chemAlF} for direct laser cooling experiments appears mainly to be limited to the group III monofluorides. Although AlCl has been produced by the analogous reaction to \eqref{eqn:chemAlF} for spectroscopic studies \cite{Hedderich1993}, AlCl$_3$ readily dimerises in the gas phase and generates a vapour pressure orders of magnitude above that of the monochloride \cite{Brunetti2009}. This rules out a similar thermochemical molecular beam for AlCl. TlF can be directly brought into the gas phase via sublimation \cite{Hunter2012,Norrgard2017}, but this molecule is much more challenging to laser cool than AlF. 

The time-averaged output of our thermochemical AlF molecular beam for the $v=0,J=1$ level is comparable to the pulsed, cryogenically cooled beam of AlF that we presented in Ref. \cite{Wright2022}. We have also shown that the thermochemical beam can be combined with a pulsed, jet-cooled supersonic AlF beam for rotationally-resolved ionisation spectroscopy using pulsed lasers, and provides access to the complementary, highly excited rotational levels. Moreover, the shot repetition rate for the AlF supersonic beam (10~Hz in this study) is ultimately limited by the vacuum load of the carrier gas. This limitation is not present with the thermochemical beam, and its data rate for pulsed laser ionisation spectroscopy could be made an order of magnitude larger with suitable laser sources.

It is worth discussing prospects for the buffer gas cooled beam in section \ref{sec:BGcooling} as a source for downstream experiments. If we couple molecules into a quadrupole electrostatic guide, weak field-seeking states with $M_J = 0, J>0$ experience a quadratic Stark shift $\delta E_{(S)}$. Within second order perturbation theory this is given as,

\begin{equation}
    \delta E_{(S)} = \frac{(\mu_X \mathcal{E})^2}{2h B_{\mathrm{rot}}} \frac{1}{(4J(J+1) - 3)} \hspace{0.5cm},
\end{equation}

\noindent with $\mu_X = 1.515(4)$~D the electric dipole moment of the $X^1\Sigma^+$ state \cite{Truppe2019}, $\mathcal{E}$ the electric field, and $B_{\mathrm{rot}}$ the rotational constant. With the electrode geometry used in Ref \cite{Junglen2004} and a peak electric field of 100~kVcm$^{-1}$, AlF molecules in the $J=2, M_J=0$ level would experience a harmonic trapping potential with a transverse velocity acceptance of $\pm12$~ms$^{-1}$, and the minimum bend radius achievable,  assuming a forward velocity of 200~ms$^{-1}$, is 30~cm. Molecules could also be delivered into a molecular Zeeman slower operating on $Q(J)$ lines of the $A^1\Pi \leftarrow{} X^1\Sigma^+$ transition. The maximum velocity change $\delta v$ across such a slower is determined by equating the change in transition Zeeman shift, $\delta E_{(Z)}/h$, to the change in the first order Doppler shift, $\delta f = \delta v/\lambda$. $\delta E_{(Z)}$ is overwhelmingly dominated by the Zeeman shift of the $A^1\Pi$ levels, which for a $\Lambda = \Omega =1, M'_J = +J'$ state is given by,

\begin{equation}
    \delta E_{(Z)} = \frac{1}{J'+1}\mu_B B \hspace{0.3cm}.
\end{equation}

\noindent  Limiting the peak magnetic field in a Zeeman slower to $B=3$~kG, readily achievable with commercial NeFeB magnets, the maximum velocity change across the slower is,

\begin{equation}
    \Delta v_{\mathrm{max}}[\mathrm{ms}^{-1}] = \frac{960}{J'+1} \hspace{0.5cm}.
\end{equation}

\noindent Hence, $\Delta v_{\mathrm{max}}$ is larger than 100~ms$^{-1}$ for $1\leq J\leq9$, and a significant fraction of the buffer gas cooled beam could be captured by such a slower. It is worth mentioning here the benefit of a transverse field Zeeman slower \cite{Ovchinnikov2007,Reinaudi2012}, which enables driving $\Delta M_J =0$ (i.e. $\pi$-polarised) transitions in the magnetic field. The branching ratio of the $\pi$-polarised decay from the $|M'_J|=J'$, $f$-symmetry levels can straightforwardly be shown to be $J'/(J'+1)$, i.e. the transverse field configuration is increasingly favoured for Zeeman slowing in excited rotational levels. 

We are not aware of any collision dynamics studies of AlF with surfaces, or of its sticking probability, although in light of the experiments in section \ref{sec:Dispenser}, these would be beneficial. Since AlF can be efficiently detected via laser-induced fluorescence in any excited rotational level of the $X^1\Sigma^+$ state, it may in fact be an interesting system to use. Our observation of room temperature AlF vapour means that loading a molecular MOT directly from a compact, cryogen-free source may be possible. Vacuum-compatible materials with a very low sticking coefficient for AlF, should they exist, would be especially useful in this context. We have built a dedicated setup to directly observe single AlF-surface collision outcomes and will report on this separately. 

\section{Conclusions}

In conclusion, we have demonstrated a continuous source of AlF molecules operating between 800 and 950~K, with a total far-field brightness of $5\times 10^{15}$~sr$^{-1}$s$^{-1}$ at 923~K, below the melting point of aluminum metal. We have applied this source in cw laser-induced fluorescence spectroscopy of the $A^1\Pi \leftarrow{}X^{1}\Sigma^+$ transition up to the $v=4$ rotational levels, and in pulsed, resonance-enhanced multiphoton ionisation spectroscopy of the $c^3\Sigma^+,v=0-8$ levels. We then implemented buffer gas cooling of the continuous beam using cold Ne, reducing the most probable velocity of the beam from 600~ms$^{-1}$ to around 200~ms$^{-1}$, and lowering its rotational temperature to near 30~K. Finally, we have demonstrated a simple molecular dispenser source, and shown that a room temperature transient AlF vapour can be generated when molecules collide with the vacuum walls of the experimental apparatus. Together, these developments will be valuable for future laser cooling experiments with AlF.

\section*{Author contributions}

Experiments were carried out by P. Kukreja, P. Agarwal, M. Doppelbauer,  S. Truppe and S. Wright with assistance from J. Cai, E. Padilla and R. Thomas; B. Sartakov and X. Liu provided theoretical support and assisted in analysing the data; H. Haak designed and constructed the Knudsen effusion cell; S. Kray and S. Wright completed other design work; S. Wright, S. Truppe and G. Meijer conceptualised the project; S. Wright and G. Meijer provided supervision; S. Wright, P. Kukreja and P. Agarwal wrote the manuscript; all authors reviewed and edited the manuscript.  

\section*{Conflicts of interest}
The authors of this study have no conflicts of interest to declare.

\section*{Data availability}

The data presented in this study will be publicly available following publication. 

\section*{Acknowledgements}

We acknowledge the expert technical assistance of the mechanical workshops of the Fritz Haber Institute. S. Wright thanks Andr\'e Fielicke, and Gert Hofstätter (AlfaVakuo e.U.), for helpful discussions.

This work was financially supported in part by the Horizon Europe project UVQuanT (Grant agreement ID 101080164).


\end{document}